\documentclass{emulateapj}        
\usepackage{color}

\usepackage{url}

\begin{document}

\slugcomment{submitted to the ApJ Supplement Series}

\title{3C\,17: The BCG of a galaxy cluster at $z=0.22$}


\author{Juan P. Madrid\altaffilmark{1}, Carlos J. Donzelli\altaffilmark{2}, Alberto Rodr\'iguez-Ardila\altaffilmark{3, 4}, 
Alessandro Paggi\altaffilmark{5, 6, 7}, Francesco Massaro\altaffilmark{5, 6, 7, 8}, Mischa Schirmer\altaffilmark{9, 10}}

\altaffiltext{1}{CSIRO, Astronomy and Space Science, PO BOX 76, Epping NSW 1710, Australia}

\altaffiltext{2}{Instituto de Astronom\'ia Te\'orica y Experimental 
IATE, CONICET - Observatorio Astron\'omico, Universidad Nacional de C\'ordoba, Laprida 854,
X5000BGR, C\'ordoba, Argentina}

\altaffiltext{3}{Laborat\'orio Nacional de Astrof\'isica/MCTIC, Rua dos Estados 
Unidos, 154, Bairro das Na\c c\~oes, Itajub\'a, MG, Brazil}

\altaffiltext{4}{Instituto de Astrof\'isica de Canarias, C/V\'ia L\'actea s/n, E-38205, La Laguna, Tenerife, Spain}

\altaffiltext{5}{Dipartimento di Fisica, Universit\`a degli Studi di Torino, via Pietro Giuria 1, I-10125 Torino, Italy}

\altaffiltext{6}{Istituto Nazionale di Fisica Nucleare, Sezione di Torino, I-10125 Torino, Italy}

\altaffiltext{7}{INAF-Osservatorio Astrofisico di Torino, via Osservatorio 20, I-10025 Pino Torinese, Italy}

\altaffiltext{8}{Consorzio Interuniversitario per la Fisica Spaziale (CIFS), via Pietro Giuria 1, I-10125, Torino, Italy}

\altaffiltext{9}{Gemini Observatory,  Southern Operations  Center, Colina El Pino s/n, La Serena, Chile}

\altaffiltext{10}{Max-Planck-Institut f\"ur Astronomie, K\"onigstuhl 17, D-69117 Heidelberg, Germany}

                         
\begin{abstract}

Gemini Multi Object Spectrograph medium-resolution spectra and photometric 
data of 39 objects in the field of the radio galaxy 3C\,17 are presented. 
Based on the new data, a previously uncataloged cluster of galaxies is identified 
at a mean redshift of $z=0.220~\pm~0.003$, a projected virial radius of 0.37 Mpc, and a 
velocity dispersion of $\sigma_v = 821~\pm$~171 km s$^{-1}$. The brightest member of this cluster 
is 3C\,17  with $M_r$ = -22.45 mag. The surface brightness profile of 3C\,17 
is best fit with two components (Exponential $+$ S\'ersic) characteristic of 
brightest cluster galaxies. The spectrum of 3C\,17 is dominated by broad emission
lines H$\alpha$~$+$~N[~II] and H$\beta$~$+$~[O~III]. Analysis of {\it Chandra} data 
shows extended emission around the cluster core that supports the existence of hot gas
cospatial with 3C\,17. The discovery of a cluster of galaxies
around 3C\,17 better explains the sharply bent morphology of the radio jet 
given that it propagates through a dense intracluster medium.

\end{abstract}


\maketitle

\section{Introduction}

The Third Cambridge Catalogue \citep[3C;][]{edge1959} and its revised version 
\citep[3CR;][]{bennett1962a, bennett1962b} contain the most powerful 
radio sources of the northern sky. As one of the earliest radio  surveys, the 
3CR was limited by the sensitivity of the telescopes at the time to nine Janskys 
as its detection limit (at 178 MHz), that is, the strongest sources of the radio luminosity 
function. 

Since its publication, the 3CR catalog has been intensively studied at different wavelengths.
The early association between radio sources and their optical counterparts was carried out
with ground-based optical imaging and spectroscopy \citep[][among many others]{wyndham1966,
kristian1974, smith1980, gunn1981, laing1983, spinrad1984, spinrad1985}. More recently, the sources 
of the 3CR catalog have received extensive follow-up in broadband imaging with the {\it Hubble 
Space Telescope (HST)} in the near-ultraviolet \citep{mccarthy1997}, optical \citep{dekoff1996,martel1999}, 
near-infrared \citep{madrid2006}, and narrow-band emission line \citep{tremblay2009}. 
{\it Spitzer} observations of 3CR sources in the mid-infrared were obtained by \citet{cleary2007} and 
\citet{leipski2010}. A study of star formation in 3CR objects was carried out by \citet{westhues2016} using {\it Herschel} data.

All extragalactic 3CR sources have now been observed in the X-rays up to $z =1.5$ through 
dedicated {\it Chandra} observing campaigns \citep{massaro2010, massaro2012, massaro2013, massaro2015,
massaro2018, balmaverde2012, wilkes2013, stuardi2018}. The 3CR catalogue has also been observed 
with {\it ROSAT} \citep{hardcastle2000}, and {\it XMM-Newton} \citep{evans2006}. 
Recent {\it Swift} observations of unidentified 3CR sources were presented by 
\citet{maselli2016}. Based on the large set of observations cited above, and many more, 
it is now clearly established that most sources in the 3CR catalog are radio galaxies or quasars.
It is also now understood that the radio emission of strong radio sources like those in the 
3CR catalog, is powered by gas accretion into supermassive black holes residing at the 
core of their host galaxies \citep{salpeter1964,lynden1969}.

Radio galaxies interact with their environment in often spectacular ways. NGC\ 1265 (3C 83.1B) is the 
prototypical example of a narrow-angle-tail radio galaxy \citep{gisler1979, odea1986}. 
In narrow-angle-tail radio galaxies, two-sided jets emanating from the central black hole 
are bent due to supersonic ram pressure of the intracluster medium \citep[ICM;][]{begelman1979}.
For NGC\ 1265, two jets emanate perpendicular to the tail, then they bend in the direction of 
the tail and finally merge \citep{sijbring1998}.

The focus of this paper is 3C\,17 (PKS~0035--02), a broad-line radio galaxy that also shows strong 
indications of interaction with its environment. The kiloparsec-scale radio morphology of 3C\,17 is dominated by a 
single-sided, dramatically curved jet described by \citet{morganti1999}. X-ray emission arising from the curved 
radio jet of 3C\,17 was discovered by \citet{massaro2009}. The ratio of X-ray to radio
intensities for the jet knots was suggested as a diagnostic tool for the
X-ray emission process, likely due to inverse Compton scattering of CMB
photons. At the same time, by combining X-ray data with {\it HST} images, an
intriguing optical object, with no radio or X-ray counterparts, was also
discovered. It was interpreted as a possible edge-on spiral galaxy
that appears to be pierced by the jet or a feature associated with the
3C\,17 jet (object 2 in Figure 1). 

Motivated by the remarkable morphology of the radio jet, and its knots, new Gemini Multi Object 
Spectrograph multislit observations are obtained with the aim of studying the environment that 
hosts 3C\,17. 

The redshift of 3C\,17 ($z = 0.22$) was determined by \citet{schmidt1965} with the 
Palomar 200 inch telescope. For this work, a flat cosmology is assumed, with 
$H_0$ = 68 km s$^{-1}$ Mpc$^{-1}$, $\Omega_{M}$ = 0.31, and $\Omega_{\Lambda}$ = 0.69
(Planck Collaboration XIII 2016). At the redshift of 3C\ 17 (i.e., $z=$0.22)
one arcsecond is equivalent to 3.65 kpc \citep{wright2006}.


\section{Gemini Observations}

Medium-resolution spectra of 3C\,17 and 38 other objects in the field are obtained using 
the Gemini Multi-Object Spectrograph (GMOS), under program GN-2016B-Q-6 (PI: C.~J. Donzelli). 
The field is centered on 3C\,17, covering a region of $\sim5\times5$ arcmin$^{2}$. 

A Gemini pre-image is obtained on 30 July 2016 and consisted of $3\times150$ s exposures 
in the Sloan $r'$ (G0326) filter; see Figures 1 and 2. The Sloan $r'$ filter has an effective 
wavelength of 630 nm. A binning of $2\times2$ pixels was used, resulting in a spatial scale 
of $0.146\arcsec$ per pixel. 

Using the pre-image above a multislit mask is created. The highest priority is given to those 
objects along the path, or near, the radio jet. Additional targets in the GMOS field are included in order 
to study the environment of 3C\,17. It should be noted that the final placement of slits on the mask is
governed by the slit-positioning algorithm (SPA). The selection carried out by the SPA is based 
on object priority and position on the frame, and preserves a minimum separation of two pixels between slits.
A total of 45 slits are placed on the mask, of which 39 are science objects. Six slits are positioned in order to obtain
sky spectra. Most slits dimensions are 1$\arcsec$ wide by 4$\arcsec$ long. However, in order to avoid 
overlapping, the slits clustered around 3C\,17 have shorter lengths. Figure 1 shows a zoomed-in view of selected 
targets near 3C\,17 labeled with the slit numbers 2, 3, 4, 5, and 9, while Figure 2 shows the remaining 
science targets across the field.

The grating in use is R400, which has a ruling density of 400 lines/mm and a spectral resolution 
of 1.44~\AA~per pixel at the blazing wavelength of 764 nm. 
Three exposures of 1200 s each are obtained with central wavelengths of 750, 760, and 770 nm that 
are used to remove the instrumental gaps between the CCDs. Spectra typically cover the wavelength range 
520-900 nm, although this range depends on the slit position.

GMOS spectra are obtained during the nights of the 5 and 22 of October 2016. Both nights 
had a seeing of  $\sim 0.65\arcsec$. Flat-fields, spectra of a standard star, 
and the copper-argon $CuAr$ lamp are also acquired in order to perform flux 
and wavelength calibrations. 


\begin{figure}
\begin{center}
\epsscale{1.25}
 \plotone{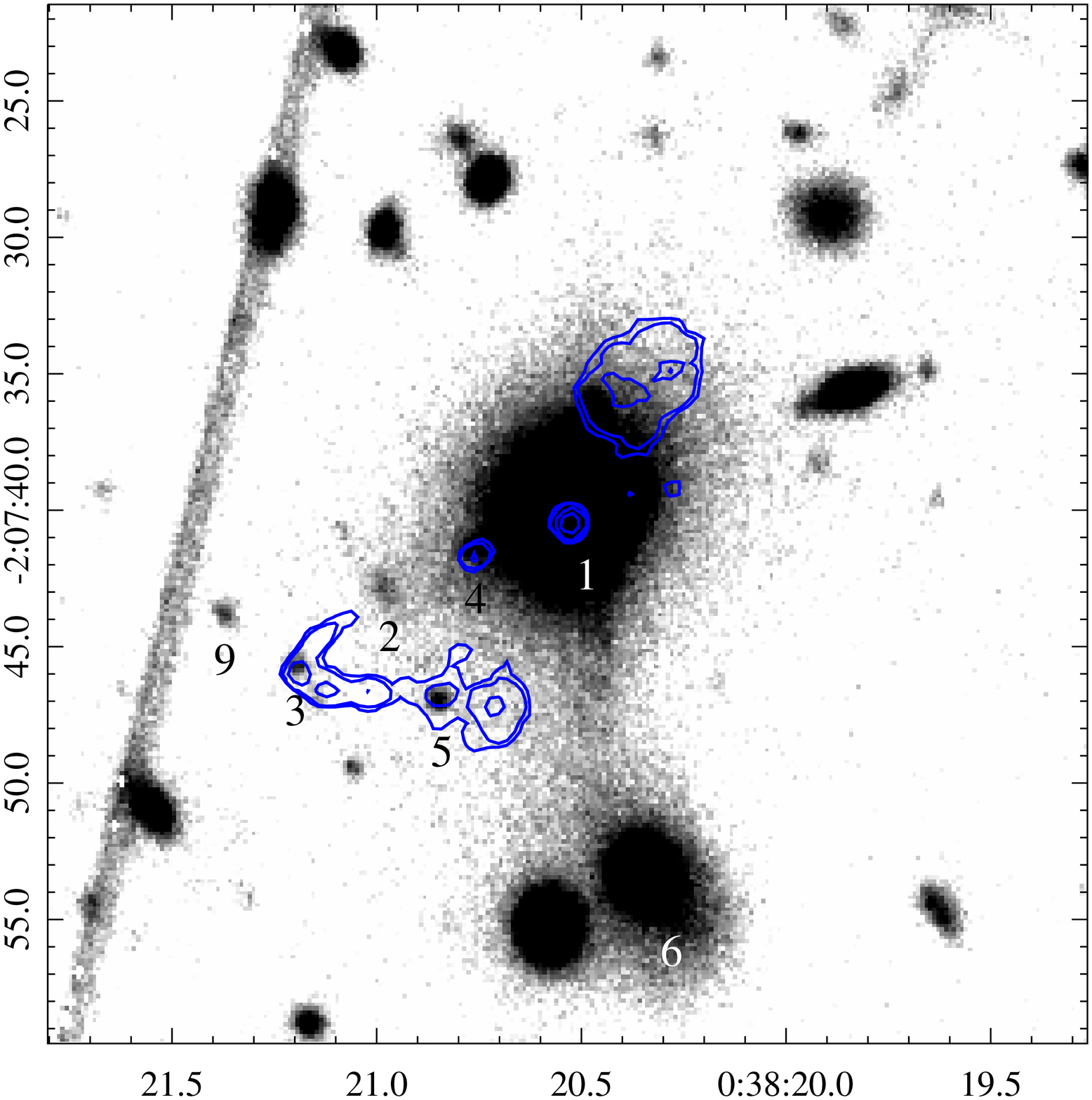}
 \caption{Central field of the Gemini pre-image. Objects are labeled with the slit number.
The image shows the central $\sim40\times40$ arcsec of the field. Blue contours show the 5 GHz VLA
map of \citet{morganti1999}. Object with label 4 is identified as S3.7 in \citet{massaro2009}.
Note that the radio jet reignites ``after" object 2, as shown in Figure 3 of \citet{massaro2009}. 
The jet sharply bends after object 3 which corresponds to object S11.3 in the \citet{massaro2009} 
nomenclature. The straight line across the image is a satellite trail, also visible in Figure 2. 
North is up and East is left.
 \label{fig1}}
 \end{center}
 \end{figure}

\begin{figure*}
\begin{center}
\epsscale{1.2}
 \plotone{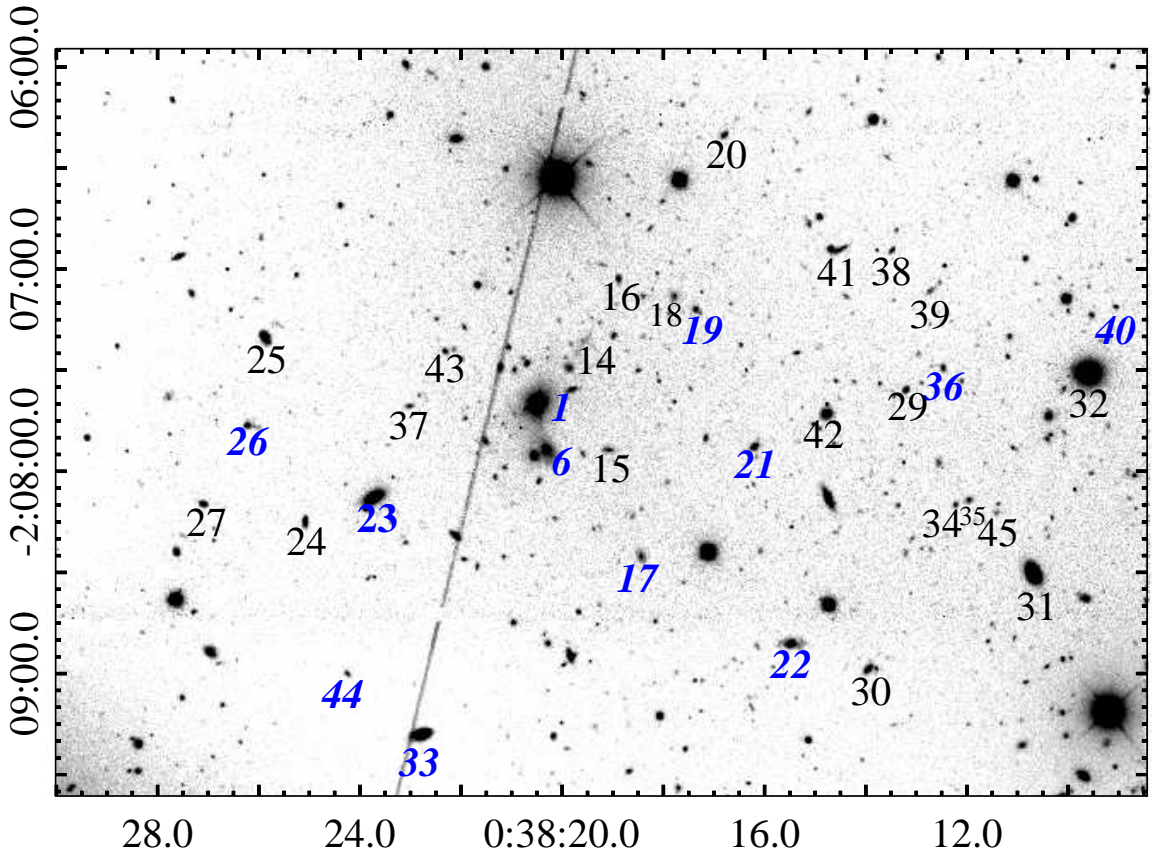}
 \caption{GMOS $r'$-band image centered on 3C 17. Galaxies selected for spectroscopy
are labeled with the slit number. Cluster members are indicated in blue, {\it italic} font.
3C 17 is labelled with number 1. North is up and east is left.
 \label{fig2}}
 \end{center} 
 \end{figure*}

\begin{center}
\begin{deluxetable*}{lccccl} 
\tabletypesize{\scriptsize} 
\tablecaption{Targets of the GMOS multislit mask \label{tbl-1}} 
\tablewidth{0pt}
\tablehead{
\colhead{Slit} & \colhead{R.A.} & \colhead{Decl.} & \colhead{z} & \colhead{m$_{r'}$} & \colhead{Comments}}
 
\startdata
1* &  00:38:20.529 &  -02:07:40.29 &  0.221 & 17.84 & 3C 17\\
2  &  00:38:20.976 &  -02:07:42.51 &  ---    & 24.39 & Linear object\\
3  &  00:38:21.197 &  -02:07:45.72 &  ---    & 24.06 & S11.3\\
4  &  00:38:20.762 &  -02:07:41.40 &  ---    & 24.47 & S3.7\\
5  &  00:38:20.847 &  -02:07:46.99 &  ---    & 24.14 & knot\\
6* &  00:38:20.351 &  -02:07:53.41 &  0.220 & 19.35 & 2nd brightest galaxy\\
7  &  00:38:19.841 &  -02:07:53.41 &  ---    & ---   & sky\\
8  &  00:38:21.586 &  -02:07:41.40 &  ---    & ---   & sky\\
9  &  00:38:21.374 &  -02:07:43.86 &  ---    & 24.15 & knot\\
10 &  00:38:20.069 &  -02:07:40.29 &  ---    & ---   & sky\\  
11 &  00:38:21.828 &  -02:07:42.51 &  ---    & ---   & sky\\
12 &  00:38:22.102 &  -02:07:45.72 &  ---    & ---   & sky\\
13 &  00:38:21.672 &  -02:07:46.99 &  ---    & ---   & sky\\
14 &  00:38:19.624 &  -02:07:21.49 &  ---    & 22.18 & Interacting\\	
15 &  00:38:19.115 &  -02:07:53.57 &  0.275 & 21.65 & Spiral\\
16 &  00:38:18.923 &  -02:07:02.83 &  0.799 & 21.71 & Irregular\\	
17*&  00:38:18.475 &  -02:08:25.22 &  0.226 & 21.32 & Spiral\\
18 &  00:38:17.814 &  -02:07:07.95 &  ---    & 21.72 & Elliptical\\
19*&  00:38:17.401 &  -02:07:11.97 &  0.222 & 21.77 & Elliptical\\
20 &  00:38:16.820 &  -02:06:20.21 &  0.670 & 21.74 & Barred\\
21*&  00:38:16.236 &  -02:07:52.66 &  0.221 & 20.71 & Elliptical\\
22*&  00:38:15.495 &  -02:08:50.91 &  0.212 & 20.30 & Spiral-Interacting\\
23*&  00:38:23.754 &  -02:08:07.66 &  0.222 & 18.48 & S0\\
24 &  00:38:25.115 &  -02:08:14.57 &  0.421 & 20.89 & Irregular\\
25 &  00:38:25.920 &  -02:07:20.17 &  0.165 & 20.54 & Spiral-Interacting\\
26*&  00:38:26.252 &  -02:07:46.41 &  0.221 & 21.16 & Elliptical-Interacting\\
27 &  00:38:27.128 &  -02:08:09.73 &  0.477 & 20.90 & Elliptical?\\
28 &  00:38:28.817 &  -02:07:22.91 &  0.160 & 22.90 & Elliptical\\
29 &  00:38:13.241 &  -02:07:35.93 &  0.410 & 21.43 & Elliptical\\
30 &  00:38:13.984 &  -02:08:58.44 &  0.255 & 20.93 & Irregular\\
31 &  00:38:10.744 &  -02:08:30.05 &  0.112 & 18.79 & Barred-Spiral\\
32 &  00:38:09.633 &  -02:07:30.86 &  0.058 & 18.02 & Spiral\\
33*&  00:38:22.805 &  -02:09:18.00 &  0.217 & 19.37 & Spiral\\
34 &  00:38:12.265 &  -02:08:09.84 &  0.398 & 22.52 & Elliptical\\
35 &  00:38:11.998 &  -02:08:08.38 &  0.402 & 22.20 & Elliptical\\
36*&  00:38:12.511 &  -02:07:29.30 &  0.220 & 22.04 & Elliptical\\
37 &  00:38:23.054 &  -02:07:40.65 &  0.321 & 22.46 & Spiral\\
38 &  00:38:13.516 &  -02:06:54.34 &  0.367 & 22.25 & Spiral\\
39 &  00:38:12.761 &  -02:07:06.50 &  ---    & 22.62 & Spiral\\
40*&  00:38:09.092 &  -02:07:11.21 &  0.217 & 22.35 & Elliptical\\
41 &  00:38:14.551 &  -02:06:54.36 &  0.671 & 21.82 & Spiral-Interacting\\
42 &  00:38:15.074 &  -02:07:45.89 &  0.463 & 23.01 & Elliptical\\
43 &  00:38:22.341 &  -02:07:24.30 &  0.320 & 22.05 & Elliptical\\
44*&  00:38:24.276 &  -02:07:59.99 &  0.218 & 22.55 & Irregular\\
45 &  00:38:11.458 &  -02:08:12.18 &  ---    & 22.98 & Irregular\\
\enddata

\tablecomments{Column 1: slit number, an asterisk denotes cluster membership. Column 2: Right Ascension (J2000). Column 3: Declination (J2000). 
Column 4: redshift. Column 5: apparent $r'$ magnitude. Column 6: Comments. While most of the spectroscopic targets 
are uncataloged the following objects have other names: slit 23: GALEXMSC J003823.80-020811.3; slit 32: GALEXASC 
J003809.60-020731.7; slit 33: GALEXMSC J003822.72-020917.6; slit 41: GALEXMSC J003823.80-020654.2}\\

\end{deluxetable*}
\end{center}

\bigskip 
\subsection{Data Reduction}

Standard data reduction is carried out with the Gemini \texttt{IRAF} package. A summary of
our data reduction procedures can be found in \citet{madrid2013}, \citet{muriel2015}, and 
\citet{rovero2016}. In the following paragraphs we describe the most important steps of 
the data reduction.

Images are pre-processed through the standard Gemini pipeline that corrects for 
bias, dark current, and flat-field. Spectral data are processed with the \texttt{gsreduce} task, which 
performs overscan, and cosmic ray removal. Bias and flat-field frames are generated with the 
\texttt{gbias} and \texttt{gsflat} routines. The sky level is removed interactively using the 
task \texttt{gskysub}, and each of the spectra are extracted with the \texttt{gsextract}
routine. Wavelength calibration is performed using the task \texttt{gswavelength}, while flux 
calibration is done with the \texttt{gscalibrate} routine, which uses the sensitivity function derived 
by the \texttt{gsstandard} task. For this purpose, the spectra of the spectrophotometric standard star $g191$ were 
taken with an identical instrument configuration. The photometric zero-point is taken from the 
Gemini website \footnote{\url{http://www.gemini.edu/sciops/instruments/gmos/calibration/photometric-stds}}. 

Redshifts are derived using the \texttt{fxcor} routine of  \texttt{IRAF} on the continuum-subtracted 
spectra. This task computes radial velocities by deriving the Fourier cross correlation between two spectra. 
The spectra for globular clusters and planetary nebulae in NGC 7793, obtained during a previous 
Gemini run, are used as templates. Redshifts are computed using four or more of the following 
lines: H$\beta$, [O~III] (4959, 5007\AA), H$\alpha$, [N~II] (6583\AA), and the Mg~I and Na absorption
lines. Target coordinates, magnitudes, and redshifts are listed in Table 1. The largest error on the redshifts 
given by the task \texttt{fxcor} is 0.001, hence the number of significant digits reported in Table 1. Photometric 
measurements are carried out using the Gemini pre-image and are explained in section \ref{photometry}.

\section{A cluster detection at the 3C\,17 redshift}

Figure 3 shows the redshift distribution of the 31 galaxies for which a redshift is measured.
There is a conspicuous cluster at $z\sim 0.2$ consisting of 12 galaxies, including 3C\,17. 
The number of cluster members should be taken as a lower threshold given that many galaxies in the field 
of view have magnitudes that are in the range of those measured for cluster members but were not 
included in the GMOS mask. Also, Figures 1 and 2 show that there is a concentration, 
or overdensity, of galaxies toward 3C\,17. Some of these galaxies show clear signs of interactions, 
such as tidal tails and plumes. There is a clear bridge between 3C\,17 and a companion galaxy to the south 
noted earlier by \citet{ramos2011}.

Figure 4 shows the Gemini spectra for the members of the 3C\,17 galaxy cluster that have emission 
lines. Figure 5 shows the spectra for cluster members identified through their absorption lines.

Table 1 contains the position and object description for all slits in the GMOS mask. It also
contains the redshift and magnitudes for those targets that have these two quantities determined. 


\begin{figure}
\epsscale{1.125}
 \plotone{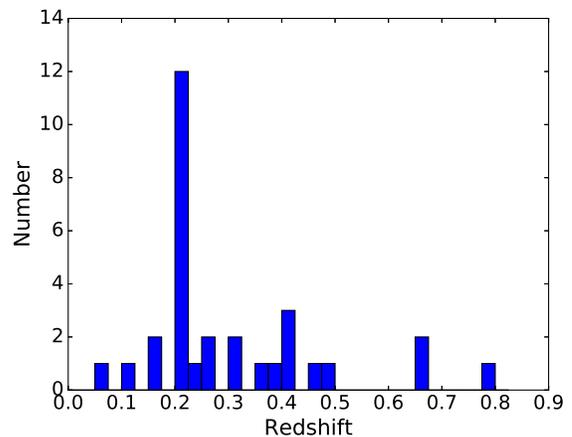}
 \caption{Redshift distribution for target galaxies. A conspicuous cluster of twelve galaxies
is present at $z$=0.2
 \label{fig3}}
 \end{figure}


The mean redshift ($z_{mean}$) and the velocity dispersion ($\sigma_v$) of this galaxy 
cluster are derived using the Gapper estimator as defined in \citet{beers1990}. An example of the
application of this method can be found in \citet{muriel2015}.

For the 3C\,17 cluster, the mean redshift is $z_{mean} = 0.220 \pm 0.003$, and its velocity dispersion
$\sigma_v = 821~\pm~171$ km s$^{-1}$. The virial radius is computed using the harmonic mean 
of the projected separations between the cluster members (see formula (1) given 
in Section 3.3 by \citet{nurmi2013}), which yields $R_{vir} = 0.37$ Mpc.
These values suggest a cluster of galaxies \citep[e.g.][]{amodeo2017}.


\section{3C\,17 Jet knots} 

The comparison between radio and X-ray emission in the curved jet of 3C\,17 was originally 
carried out by \citet{massaro2009} to highlight jet knots with X-ray counterparts.
Two of these X-ray knots, S3.7 and S11.3, are particularly interesting because they 
also have optical counterparts (which correspond in Fig.\ 1 to 
slits 4 and 3, respectively). Even more intriguing is the fact that the jet 
sharply bends after one of these knots, S11.3 (slit 3 in Figure 1). Some of these knots are 
also present in the near infrared image of \citet{inskip2010}.

Slits are placed on each of these aforementioned knots in an effort to determine their nature.
Figure 1 shows a zoomed-in view of the central region around 3C\,17 with objects selected 
for spectroscopy. Emission lines are not detected for knots 2, 3, 4, 5, and 9. The Gemini data 
place an upper limit of $10^{-18}$ erg cm$^{-2}$ s$^{-1}$ \AA$^{-1}$ for the flux of any 
potential emission line present on the jet knots described above.

The absence of detectable emission lines for slit 2, in particular, is relevant. The object labeled 2 
in Figure 1 is roughly perpendicular, in projection,  to the path of the jet; see also the {\it HST} image 
presented by \citet{massaro2009}. A plausible explanation for the nature of this object is an emitting 
region that arises from the interaction of the jet with a gas cloud (e.\ g.\ {\sc H\,i}), which is 
shock-ionized by the jet and produces free-free emission \citep{massaro2009}. The absence of emission 
lines weakens the possible presence of shocks.

\begin{figure}
\epsscale{1.3}
 \plotone{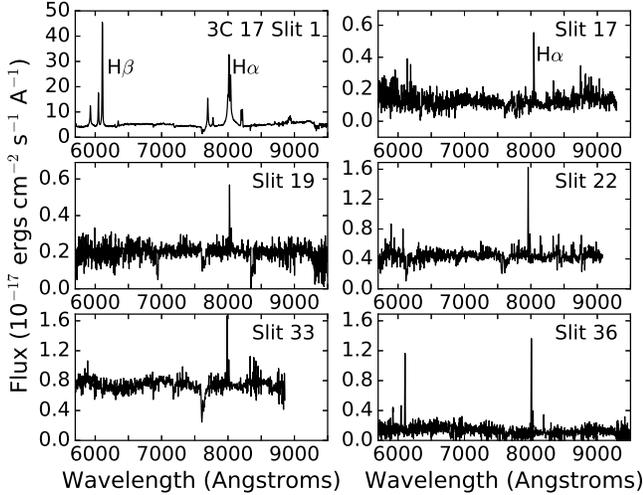}
 \caption{Spectra for the cluster galaxies with emission lines. These spectra
 are shown at their observed wavelength; that is, they are not corrected for
 radial velocities. The prominent emission line at 8006~\AA~is H$\alpha$.
 \label{fig4}}
 \end{figure}


\begin{figure}
\epsscale{1.3}
 \plotone{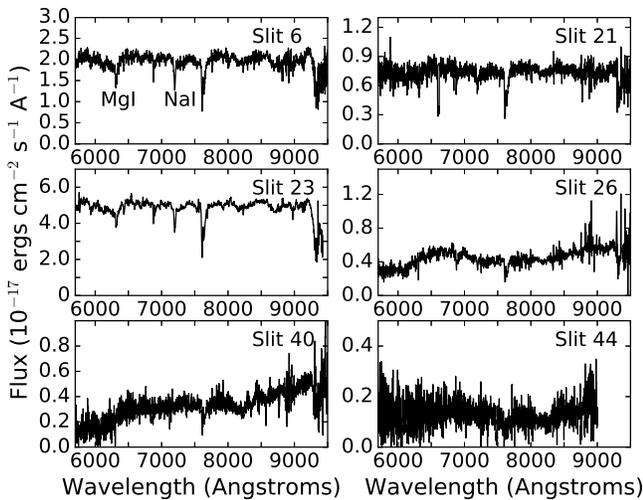}
 \caption{Spectra for the cluster galaxies identified by their absorption lines.
 These spectra are not corrected for radial velocities. The prominent absorption
 feature at 7610~\AA~is a telluric line. 
 \label{fig5}}
 \end{figure}


\section{The Spectrum of 3C\,17}\label{3c17spectrum}


The optical spectrum of 3C\,17 is also obtained with the new Gemini data. An analysis
of this spectrum is given in this section. In order to study the physics of the ionized 
gas in the 3C\,17 core, it is first necessary to correct the data for Galactic 
extinction. Here, a value of E(B-V) = 0.023 mag is adopted, as determined by \citet{schlafy2011}. 
The presence of H$\alpha$ and H$\beta$ also allow us to assess the extinction affecting the
Narrow-Line Region (NLR). For this purpose, an intrinsic Balmer decrement of 3.1, typical of
AGNs \citep{osterbrock1989}, is assumed. The observed narrow emission-line flux ratio 
of H$\alpha$/H$\beta$ = 3.15 indicates that the amount of dust in the line of sight of the NLR 
is negligible.


\begin{figure}
 \plotone{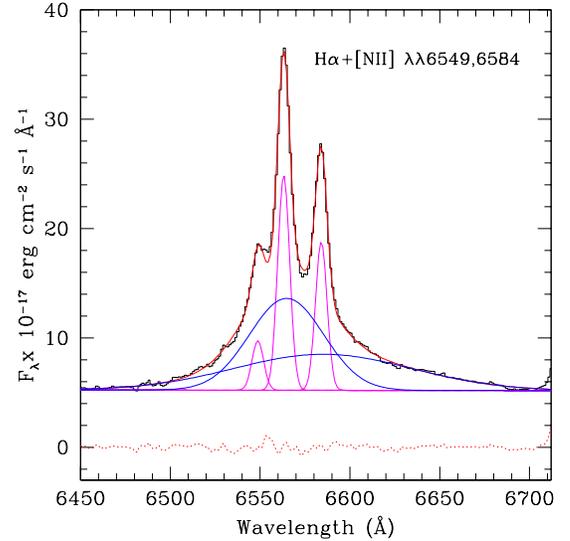}
 \caption{Example of the deblending procedure applied to H$\alpha$. 
The black histogram line is the observed spectrum, and the 
total fit is in red. The dashed line is the residual after subtraction of the
fit. Components associated with the NLR are in magenta, while those associated with
the BLR are in blue. Note that the redshifted H$\alpha$ emission is observed at 8006~\AA; 
see Figure 1.
 \label{fig7}}
 \end{figure}


\begin{figure}
 \plotone{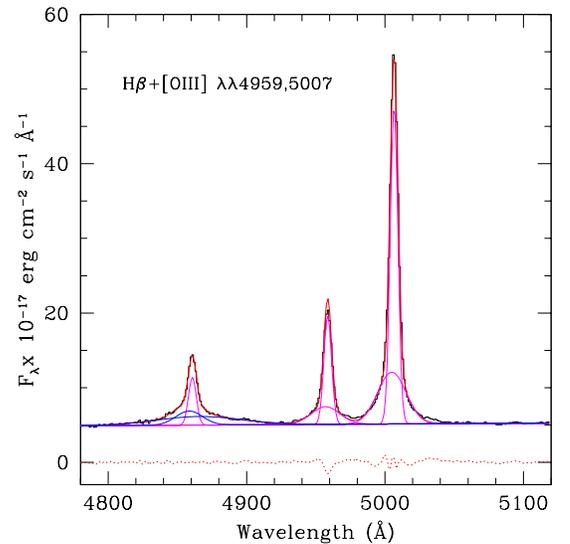}
 \caption{Same as Fig.\ 6 but for H$\beta$.
 \label{fig8}}
 \end{figure}


The optical spectrum extracted from the nuclear region of 3C\,17
is dominated by prominent emission lines of H$\alpha$+[N\,{\sc ii}],
H$\beta$, and [O\,{\sc iii}] on top of a continuum with evidence of stellar
absorption features. The base of the Balmer lines shows the presence
of broad components, confirming that this source is a broad-line object
(BLO) as suggested by \citet{baldi2013}.
Low-ionization narrow emission lines of [S\,{\sc ii}]~$\lambda\lambda$6717, 6731, and
[O\,{\sc i}]~$\lambda\lambda$6300, 6363 are also conspicuous in the
spectrum. The redshift derived from the position of these lines is
$z=0.220 \pm 0.001$, in very good agreement with recent values found in the literature
for this object \citep{buttiglione2010, baldi2013}.

Emission-line fluxes on the 3C\,17 spectrum are measured using the LINER routine 
\citep{pogge1993}, a $\chi$-squared minimization algorithm that 
can fit simultaneously up to eight profile functions to a given line 
or set of blended lines. In all cases, the underlying continuum was
fit by a low-order polynomial. LINER also provides values for 
the peak position and the full width at half-maximum (FWHM) of each profile 
function. The number of Gaussian components fit varied depending on
the emission line. In order to deblend the broad and narrow components 
of H$\alpha$ and H$\beta$, three Gaussian profiles were necessary for each 
line (see below). Figures 6 and 7 show examples of the deblending procedure for 
the spectral regions containing H$\alpha$ and H$\beta$. Column 4 of Table 2 lists 
the emission-line fluxes measured in the spectrum of 3C 17. 


\begin{deluxetable}{lccc}
\tablecaption{3C 17 Emission Lines\label{tbl-2}} 
\tablehead{
\colhead{Line} & \colhead{Wavelength} & \colhead{FWHM} & \colhead{Flux}\\
\colhead{} & \colhead{(\AA)} & \colhead{(km s$^{-1}$)} & \colhead{($10^{-17}$ erg s$^{-1}$\,cm$^{-2}$)}
}

H$\alpha_n$	   &   6563.1	&   351	&   164\\
H$\alpha_b$	   &   6564.8	&  2223	&   435\\
H$\alpha_{vb}$  &   6585.2	&  5326	&   409\\
$[$N~II$]$	   &   6584.1	&   341	&   110\\
$[$S~II$]$		   &   6717.1	&   366	&    61\\
$[$S~II$]$	   	   &   6731.5	&   365	&    64\\
$[$O~I$]$$_n$   	   &   6300.4	&   356	&    70\\
$[$O~I$]$$_i$   	   &   6301.8	&  1016	&    87\\
H$\beta_n$	   &   4860.6	&   430	&    52\\
H$\beta_b$	   &   4861.6	&  2218	&    75\\
H$\beta_{vb}$	   &   4876.6	&  5225	&    61\\
$[$O~III$]$$_n$ 	   &   5006.6	&   381	&   296\\
$[$O~III$]$$_i$	   &   5004.9	&  1435	&   178\\
HeII		   &   4684.0	&   440	&    10\\
\tablecomments{The subscript $n$ stands for narrow component; likewise, $i$ stands for intermediate;
$b$ for broad; and $vb$ is for very broad. }

\end{deluxetable}


Column~3 of Table 2 displays the values of FWHM found from the fit. 
These values are already corrected in quadrature by an instrumental broadening of 73~km\,s$^{-1}$, 
measured from the sky lines present in the spectra and CuAr frames obtained 
for wavelength calibration.

The narrow component of H$\alpha$ has a FWHM of 351$\pm$10~km\,s$^{-1}$ in velocity 
space, compatible with an origin in the NLR of 3C\,17. 
The forbidden lines of [N\,{\sc ii}], [S\,{\sc ii}], [O\,{\sc i}], and [O\,{\sc iii}] 
have similar FWHMs. Note that in the latter two lines we also found the presence of 
an intermediate component, associated with the NLR, with FWHMs of 1016~km\,s$^{-1}$ and 1435~km\,s$^{-1}$, 
respectively. These lines are labeled with the subscript $i$ in Table 2. In the case of [O\,{\sc iii}], 
the centroid of that component is blueshifted by 120~km\,s$^{-1}$ relative to that of the narrower component. 

We associate this blueshifted [O\,{\sc iii}] emission with outflowing gas. Indeed, blue asymmetries in 
this line are usually indicative of nuclear outflows \citep{komossa2008,zakamska2016,marziani2016}. 
The fact that the slit in 3C\,17 was positioned along the radio jet supports 
this scenario. The presence of a broader component in [O\,{\sc i}] is most likely due to a strong 
blend with [S\,{\sc iii}]~$\lambda$6312, as such a component is not present in [O\,{\sc i}]~$\lambda$6363.
[S\,{\sc iii}]~$\lambda$6312  is a forbidden line, observed in narrow line regions and used to determine
gas temperature \citep{contini1992}.

The presence of a broad-line region in 3C\,17 is clearly manifested by
the detection of two broad components in the Balmer lines with FWHMs of 2220$\pm$30 km\,s$^{-1}$ 
and 5300$\pm$90~km\,$^{-1}$ (marked in Table 2 with the subscripts $b$ and $vb$, for broad and very broad 
components, respectively). While these peaks are very close in position with the rest-wavelength 
expected for H$\alpha$ and H$\beta$ ($\leq$ 2 \AA), the latter is significantly
redshifted, with the peak located $\sim$1000~km\,s$^{-1}$ relative to 
the narrow component of that line. Hints of broad wings for the H$\beta$ line were reported earlier by
\citet{tadhunter1993}.

In optical spectroscopy, host galaxies with broad emission lines significantly Doppler-shifted 
from the corresponding narrow lines  have been interpreted as the consequence of a late-stage 
supermassive black hole (SMBH) merger, or, alternatively, due to a kick received by the broad-lined 
SMBH resulting from anisotropic gravitational emission \citep{popovic2012, komossa2012}. In the case of 3C~17, 
the velocity shift of the very broad component and the narrow component amounts to $\sim$1000~km\,s$^{-1}$. 
This value is well within the expected range of velocity separation between the narrow and 
broad components  predicted in both scenarios above \citep{campanelli2007}. 


\section{Photometric measurements}\label{photometry}

Total magnitudes for the 39 spectroscopic targets are derived using aperture photometry
on the Gemini pre-image. Photometric measurements are obtained with the task \texttt{phot} within 
\texttt{IRAF} \texttt{daophot}. Total magnitudes are obtained by fitting a series of concentric 
circular diaphragms until the total flux converged. Sky fitting and background subtraction are 
carried out around each object, attentively avoiding stars and neighboring objects. The magnitude for
3C\,17 is derived by fitting a surface brightness profile as described below.
Total magnitudes are listed in Table 1. Considering that the mean 
redshift to the cluster is $z = 0.220$, the brightest cluster galaxy (BCG) 3C\,17 has an absolute 
magnitude of M$_r$ = -22.45 mag. The second brightest galaxy corresponds to slit 23 and it has an 
absolute magnitude of M$_r$ = -21.52 mag. 



\begin{figure}
\epsscale{1.2}
 \plotone{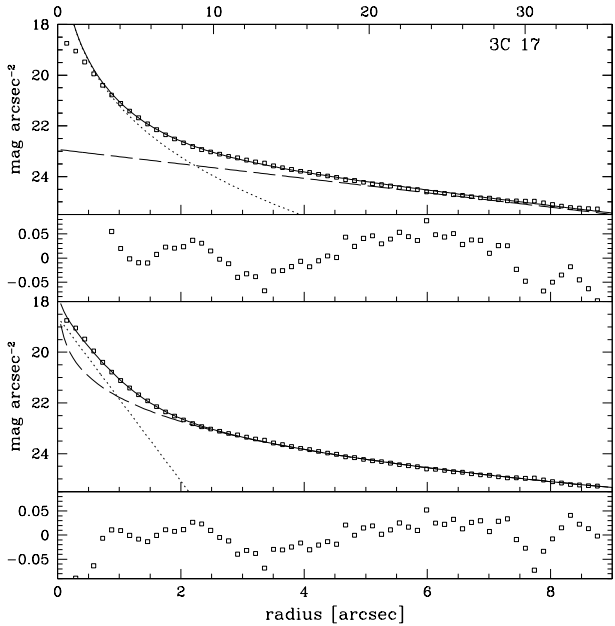}
 \caption{Luminosity profile of 3C 17. Upper panel shows the S\'ersic + exponential fit, while
the lower pannel shows the exponential + S\'ersic fit.
 \label{fig12}}
 \end{figure}


\begin{deluxetable*}{lcccccccc} 
\tabletypesize{\scriptsize} 
\tablecaption{Parameters of the 3C 17 Surface Brightness Profile\label{tbl-4}} 
\tablewidth{0pt}
\tablehead{
\colhead{Obj. + fit} & \colhead{$\mu_e$} & \colhead{$r_e$} & \colhead{$n$} & 
\colhead{$\mu_0$} & \colhead{$r_0$} & \colhead{$M_{r'}$} & \colhead { chi-sq } & \colhead { rms } \\
\colhead{ } & \colhead{ } & \colhead{ kpc } & \colhead{ } & \colhead{ } 
& \colhead{ kpc } & \colhead { } & \colhead { } & \colhead { }
} 

\startdata
3c 17 (S+E)  & 18.06 & 0.9  & 5.04 & 22.93 & 13.9 & -22.45 & 21 & 9 \\
3c 17 (E+S)  & 25.76 & 40.5 & 4.90 & 18.61 & 1.2  & ...    & 5  & 4 \\
\enddata

\tablecomments{Column (1): object and fit type, S\'ersic plus exponential (E+S), or 
exponential plus S\'ersic (E+S);  column (2): effective surface magnitude;  column (3):
effective radius (kpc); column  (4): $n$ parameter;
column (5): central surface magnitude; column (6): scale length (kpc); column (7):
Absolute r' magnitude; column (8): chi-sq parameter; column (9): rms.}

\end{deluxetable*}


\subsection{The Luminosity profile of 3C\,17}

The luminosity profile for 3C\,17 is analyzed following the methods described in 
\citet{donzelli2011} and \citet{madrid2016}. In summary, the galaxy light profile is extracted using the 
{\sc ellipse} routine \citep{jed1987} of the Space Telescope Science Analysis System
(STSDAS) package. For 3C\,17, galaxy overlapping is an issue that we circumvent by applying the technique 
described in \citet{coenda2005}. This technique consists of masking all overlapping galaxies 
before profile extraction and constructing a model of the target galaxy using its luminosity profile. 
This model is then subtracted from the original image and the resulting residual image is used
to isolate the profile of overlapping galaxies. We iterate this process until the profile of the target 
galaxy converges.

The S\'ersic profile \citep{sersic1968},

\begin{equation} \label{Ser}
I(r)=I_{e} \exp\Big\{-b_n\Big[\Big(\frac{r}r_{e}\Big)^{\beta} - 1\Big]\Big\}
\end{equation}

is used for surface brightness profile fitting. In the expression of the S\'ersic profile given above,
$I_{e}$ is the intensity at $r = r_{e}$, the radius that encloses half of the total luminosity, and
$b_n$ can be calculated using $b_n \sim 2\/n-0.33$, and $\beta = 1/n$ \citep{caon1993}. 

Profile fitting is done with the \texttt{nfit1d} routine within STSDAS. This task uses 
a $\chi^2$ minimization scheme to fit the best nonlinear functions to the galaxy 
light profile. Isophote fitting is performed down to a count level of 2$\sigma_{sky}$ that 
corresponds to a surface magnitude of $\mu_{r} \sim$25.5 mag/arcsec$^{-2}$.

For 3C 17 a single S\'ersic model fails to properly fit the luminosity profile. It is then 
necessary to add an outer exponential function \citep{freeman1970}:

\begin{equation} \label{freeman}
I(r)= I_{0} \exp\Big(-\frac{r}r_{0}\Big)
\end{equation}

In the equation above, $I_{0}$ is the central intensity  and $r_{0}$ is the scale length. 
As suggested by \citet{donzelli2011}, this exponential component is the simplest 
function to account for the `extra-light' observed in many Brightest Cluster Galaxies galaxies. 
The necessity of including an outer component on the surface brightness modeling is due to the extended 
envelopes common of BCGs. The extra exponential required to fit the brightness profile of 3C\,17
suggests that this galaxy is a typical BCG. The most accurate fit for the light profiles of BCGs is indeed 
a two component model, with an inner and outer component \citep[e.g.][]{gonzalez2005,seigar2007,donzelli2011}.
The fact that the host galaxy of 3C\,17 is a BCG gives further support to the binary 
SMBH merger hypothesis, discussed in Section \ref{3c17spectrum}, as the likely evolutionary 
scenario to give rise to broad emission lines Doppler-shifted from the corresponding narrow lines.
Corroboration of this hypothesis would require long-term monitoring.

Figure 8 shows the luminosity profile for 3C\,17 together with the fitting functions.
The upper panels show the classic S\'ersic plus exponential fit, while the lower panels show
the exponential plus S\'ersic fit. Fitting parameters are shown in Table 3.
Taking a look at the chi-squared and rms coefficients from this table we can also compare the quality 
of both types of fit. It is clear that, at least from a mathematical point of view, the inner exponential 
plus S\'ersic fit gives a better fit to the luminosity profile than the classical S\'ersic $+$ 
outer exponential.

\section{X-Ray data}
\begin{figure*}
        \centering
        \label{tab:example_table}
        \begin{tabular}{cc}  
                \includegraphics[scale=0.4]{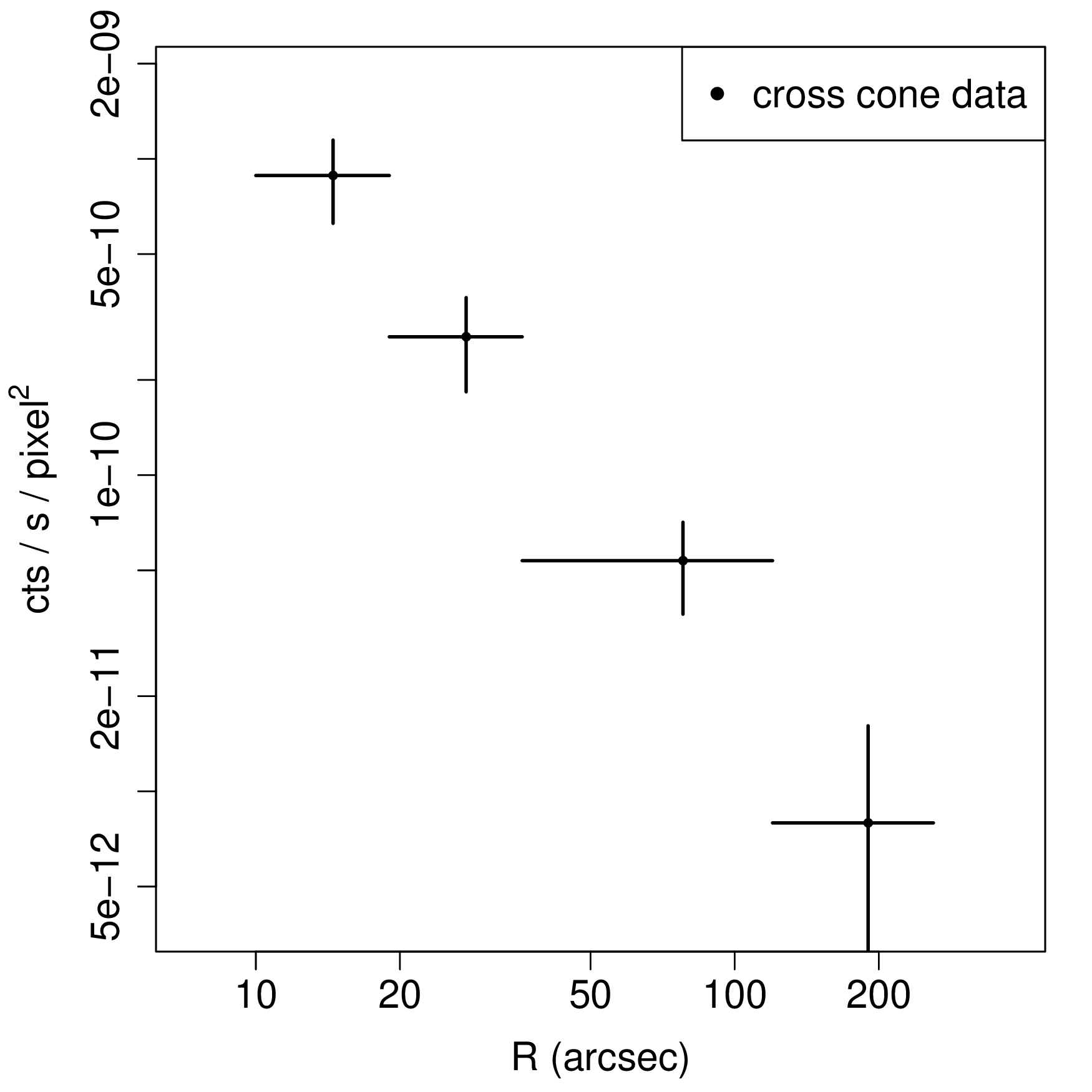} & \includegraphics[scale=0.4]{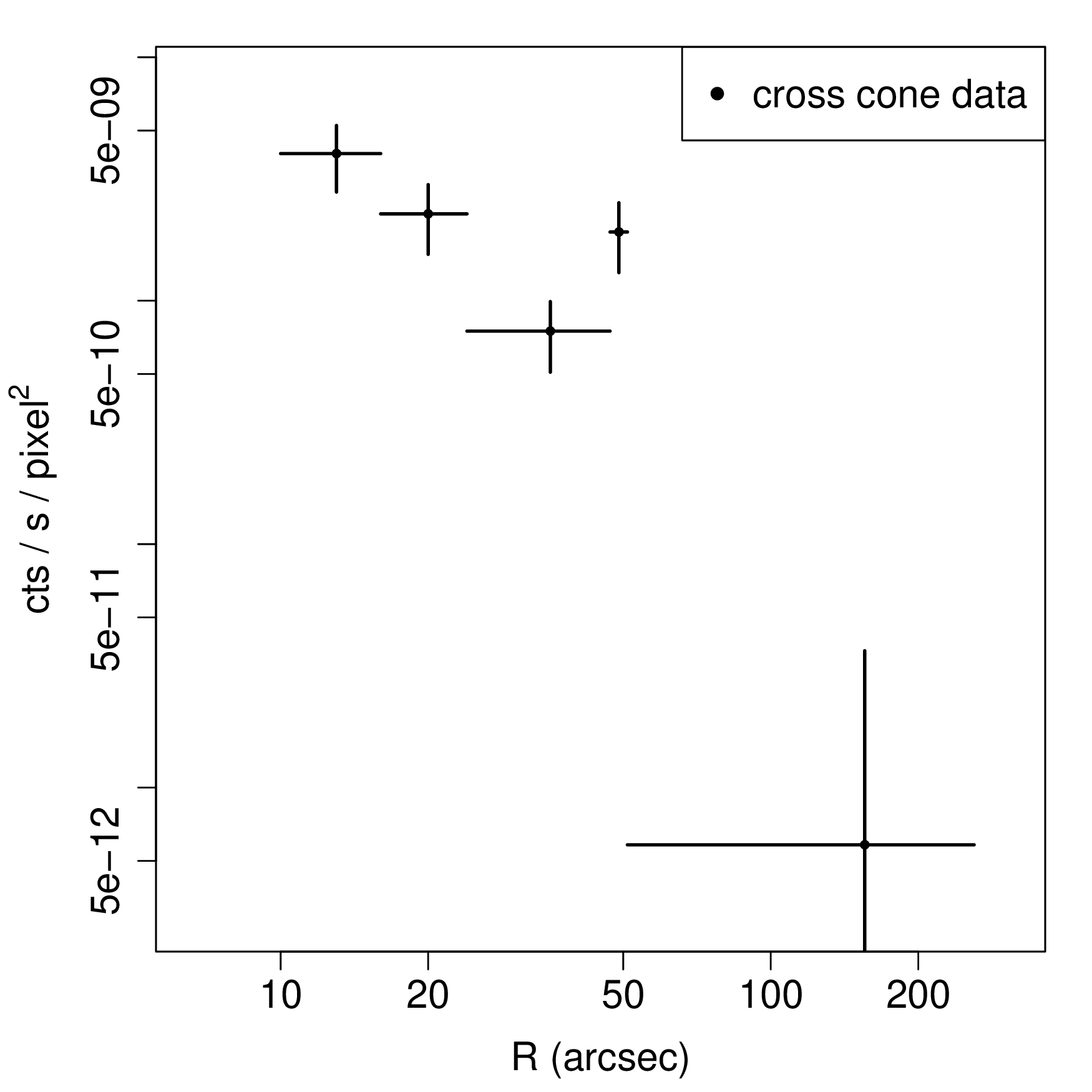}\\
        \end{tabular}
    \caption{Background-subtracted, exposure-weighted surface brightness profiles in the \(0.5-2\mbox{ keV}\) band 
(left panel) and \(0.5-7\mbox{ keV}\) band (right panel) extracted in the NE-SW direction (see the main text). 
The widths of the bins are adaptively chosen to reach a minimum signal-to-noise ratio of 3.\\
}
\end{figure*}

\begin{figure*}
        \centering
        \label{tab:example_table}
        \begin{tabular}{cc}  
                \includegraphics[scale=0.454]{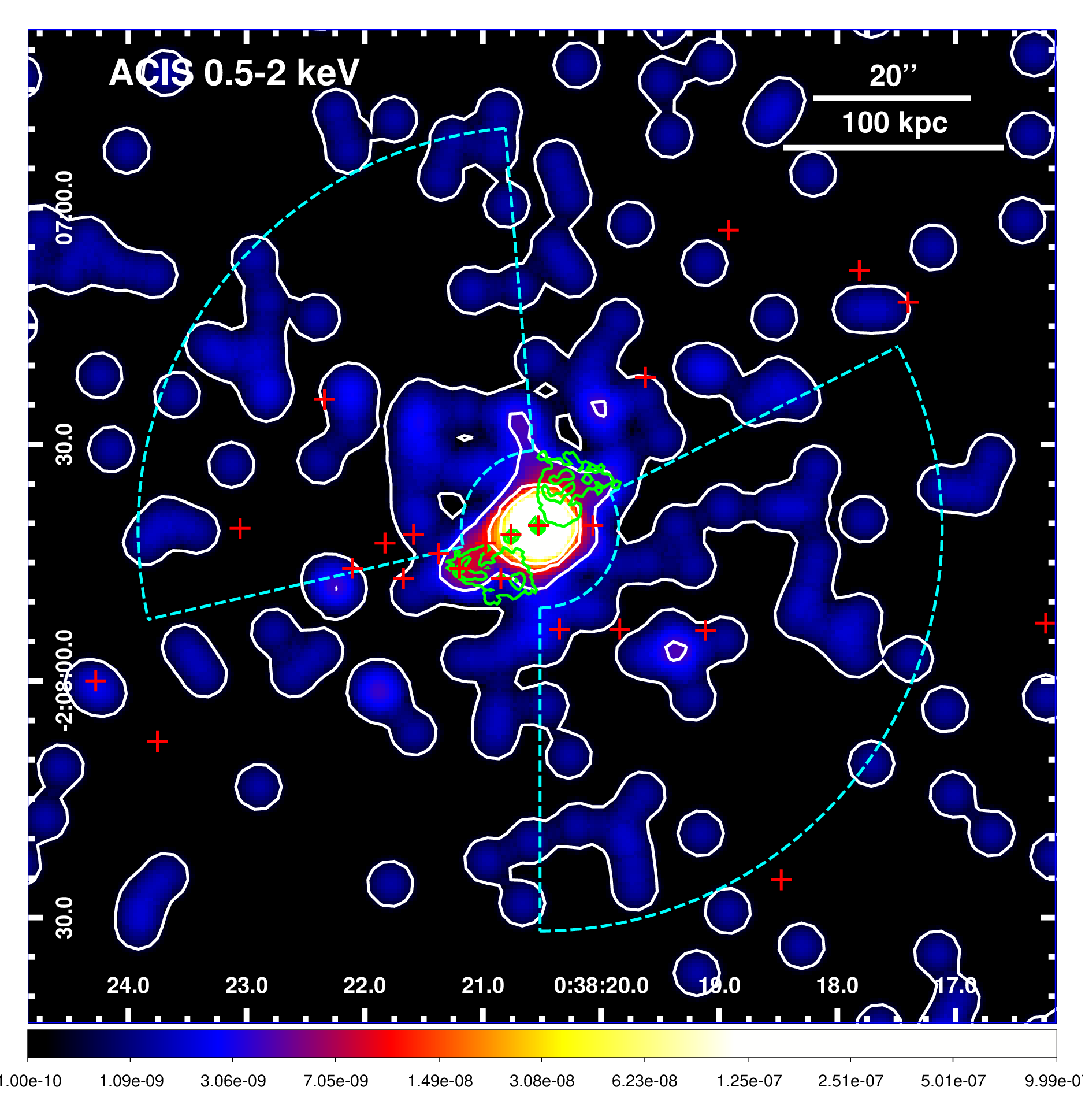} & \includegraphics[scale=0.441]{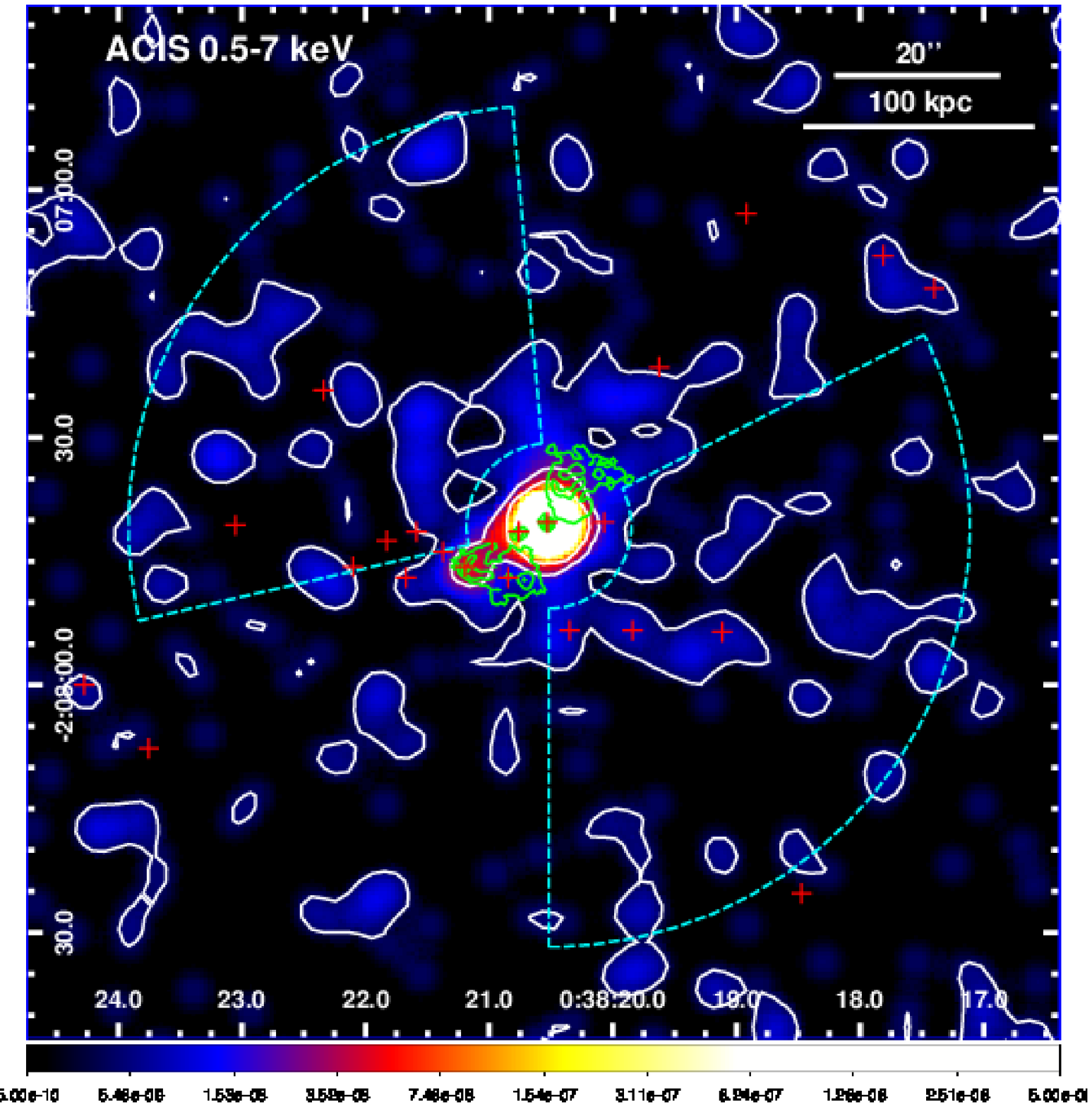}\\
        \end{tabular}
    \caption{Broadband \(0.5-2\text{ keV}\) (left panel) and \(0.5-7\text{ keV}\) (right panel) 
ACIS-S flux images centered on 3C\,17. These two images are smoothed  with a $8\times8$ pixel ($3.94\arcsec\times3.94\arcsec$) 
gaussian kernel. The green contours are from the 5 GHz VLA map of Morganti et al.\ (1999) with a restoring 
beam of $0.\arcsec4\times0.\arcsec 4$, the white lines are the contours of the X-ray flux image, and the red crosses 
indicate the galaxies in the field. The cyan dashed cones are the regions used for the extraction of 
the surface brightness profiles, with the last bin where the signal-to-noise ratio is at least three extending 
up to $\sim$50$\arcsec$.\\
}
\end{figure*}


\begin{figure}
\epsscale{1.2}
\plotone{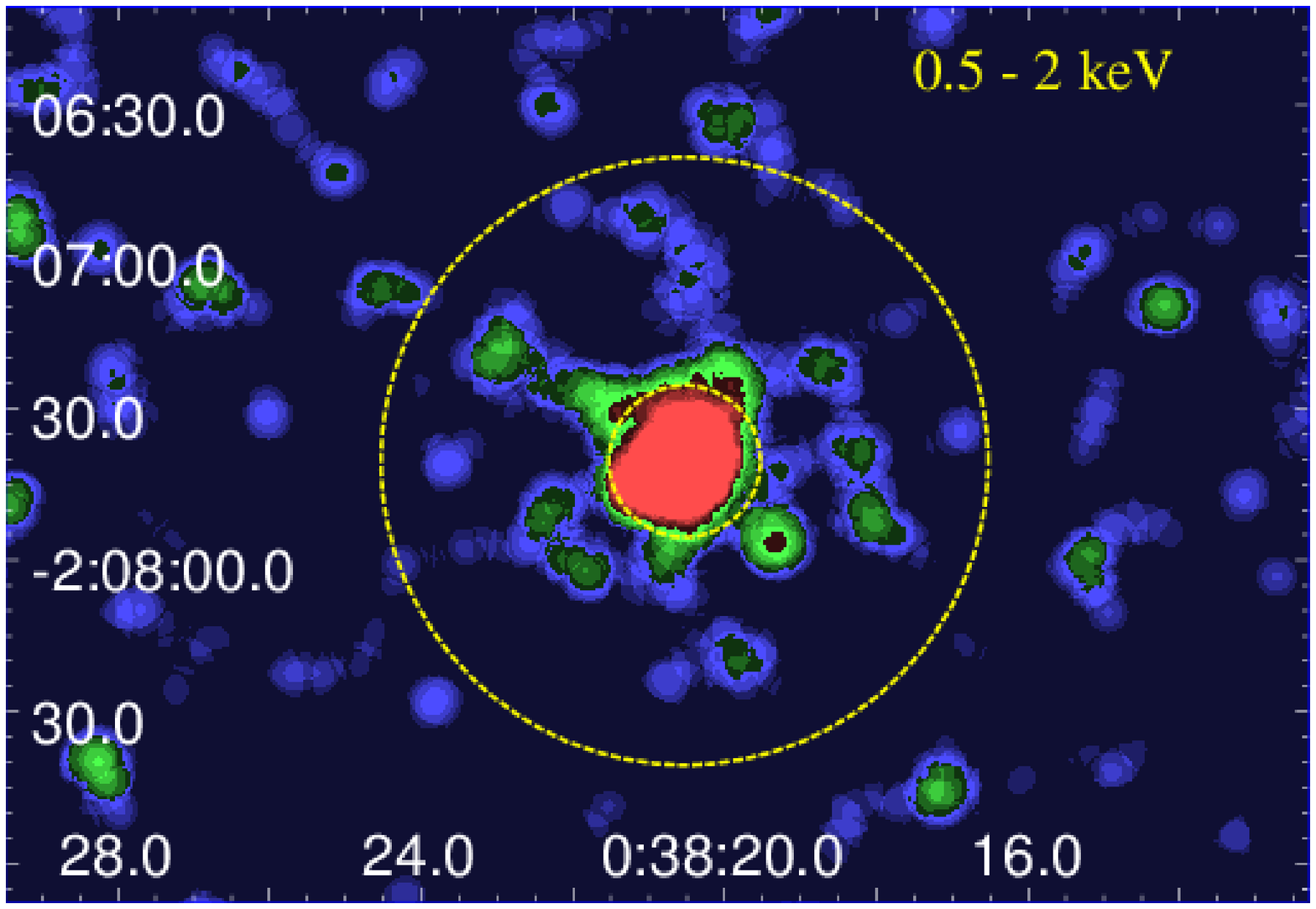}
\caption{Chandra broadband \(0.5-2\text{ keV}\) image smoothed with a Gaussian kernel of 16 pixels.
This image shows the annulus (dotted yellow lines) where the emission emanating from the ICM is 
measured. Note that this image has a bigger field of view than the images shown in Figure 10.
 \label{chandragaussian}}
\end{figure}


\textit{Chandra} data are retrieved from the Chandra Data Archive through the ChaSeR 
service\footnote{http://cda.harvard.edu/chaser}. Archival data consist of a \(8.0\) ks 
observation (9292, PI: Harris, GO 9) performed in February 2008 in VFAINT mode. These data 
are analyzed with the \textsc{CIAO} data analysis system \citep[version 4.9;][]{fruscione2006} 
and \textit{Chandra} calibration database CALDB version 4.7.6, adopting standard procedures.

Figure 9 features two broadband,  \(0.5-2\) and \(0.5-7\) keV  \textit{Chandra} images centered on ACIS-S chip 7. 
Point-spread function (PSF), exposure, and flux maps are produced using the task \textsc{mkpsfmap}.
PSF and exposure maps are used to detect point sources in the \(0.5-7\) keV energy band with the 
\textsc{wavdetect} tool, adopting a \(\sqrt{2}\) sequence of wavelet scales (i.e., 1, 2, 4, 8, 16,
and 32 pixels) and a false-positive probability threshold of \({10}^{-6}\).

To estimate the background emission we use blank-sky data included in the \textit{Chandra} 
CALDB and re-projected to the same tangent plane of the observation. We then 
produce background event files for each ACIS chip, and rescale them to match the observation
count rate in the \(9-12\mbox{ keV}\) band. 

To evaluate the physical scale of the X-ray emission we extract background-subtracted, exposure-weighted 
surface brightness profiles from the source-removed event files, shown in Fig.\ 10. The widths of the bins 
are adaptively determined to reach a minimum signal-to-noise ratio of three. For this analysis we choose 
the directions between PAs 15$^{\circ}$  and 103$^{\circ}$, and between 180$^{\circ}$  and 297$^{\circ}$, 
in order to avoid the emission observed in the radio in the GHz range. We also exclude the inner (10$\arcsec$) 
since this observation is affected by severe pile-up, as shown in the map produced with the \textsc{pileup\_map} tool. 

We find that in the \(0.5-7\) keV band the last bin that reaches a signal-to-noise ratio of 3 extends up to \(51\arcsec\). 
This extended X-ray emission, mainly in the soft  \(0.5-2\text{ keV}\) band, could be due to the presence of hot gas 
in the intergalactic medium. We measure the X-ray flux in an annulus between 5 arcsec (inner radius) and 60 arcsec 
(outer radius) from the nucleus of 3C\,17; we thus exclude the contribution from the AGN, as shown in Figure 11. 
The flux measured for this annulus is $3.84 \pm 2.55 \times 10^{-14}$ in cgs units ($10^{-15}$ erg cm$^{-2}$ s$^{-1}$). 
This is the integrated flux in the 0.5-2 keV energy range, where most of the diffuse X-ray emission due to the 
ICM is expected. The X-ray flux above yields an ICM X-ray luminosity of $L_X= 5.77 \pm 3.83\ \times10^{42}$
erg/s. 

The reason why the 3C\,17 jet sharply bends is better understood with the identification of a cluster 
of galaxies. Indeed, the jet emitted by 3C\,17 propagates through a dense intracluster medium. 
Bent-tail galaxies, or radio-loud AGN with jets that bend as they propagate through the dense ICM, 
have been reported  at high redshift. For instance, \citet{blanton2003} discovered a cluster of 
galaxies, at $z$=0.96, using a combination of VLA imaging and Keck spectroscopy. 3C\,17 also has an 
extended radio emission detected at 5 GHz with the VLA \citep{morganti1993}
although it appears to have a smaller spatial scale than the diffuse X-ray emission. 


\section{Final remarks}

An optical and X-ray study of  3C\,17 and its environment was carried out using 
Gemini and {\it Chandra} data. The presence of a strikingly curved radio jet and several
unclassified sources along the path of the jet are principal motivations for this
study. The main findings of this work are as follows:

\begin{itemize}

\item 3C\,17 is part of a galaxy cluster at a redshift of $z=0.220 \pm 0.003$.
Twelve cluster members have their redshifts determined with Gemini spectra.

\item The velocity dispersion of these 12 members is $\sigma_v = 821~\pm~171$ km s$^{-1}$,
consistent with the velocity dispersion of a cluster of galaxies.

\item The spectrum of 3C\,17 is dominated by broad emission lines.

\item 3C\,17 is the brightest cluster member. The 3C\,17 surface brightness profile
is best fit with a double component model characteristic of BCGs.

\item The analysis of the surface brightness profiles of archival {\it Chandra} data 
reveals the presence of extended soft X-ray emission surrounding 3C 17 that is likely
to originate in the hot gas of the intergalactic medium.

\end{itemize}
\bigskip

Properly characterizing the environment of AGNs with bent jets is relevant given 
that these radio sources are often used as tracers of high-redshift clusters
\citep[e.g.][]{blanton2000, wing2011, paterno2017}. Bent-tail radio galaxies will also be used 
as signposts for distant galaxy clusters in future wide-field radio-continuum surveys, 
e.\ g.\ the Evolutionary Map of the Universe (EMU) to be carried out with the Australian 
Square Kilometer Array Pathfinder \citep{norris2013}.\\


\acknowledgments

This work was the original idea of our late friend and colleague 
Dan E.\ Harris. Dan passed away on 2015 December 6th. We dedicate this
work to his memory. Dan's career spanned much of the history of radio 
and X-ray astronomy. His passion, insight, and contributions will 
always be remembered. 

We are grateful to Ron Ekers and Elizabeth Mahony for sharing their expertise
on radio galaxies with us. The VLA map shown in Figure 1 was kindly provided
by Raffaella Morganti.

This work is based  on observations  obtained at  the Gemini  Observatory,  which is
operated by the Association of Universities for Research in Astronomy,
Inc.,  under a cooperative  agreement with  the NSF  on behalf  of the
Gemini partnership:  the National Science  Foundation (United States),
the   National  Research  Council   (Canada),  CONICYT   (Chile),  the
Australian    Research   Council   (Australia),    Minist\'{e}rio   da
Ci\^{e}ncia, Tecnologia e  Inova\c{c}\~{a}o (Brazil) and Ministerio de
Ciencia, Tecnolog\'{i}a e Innovaci\'{o}n Productiva (Argentina).

The authors wish to recognize and acknowledge the very significant 
cultural role and reverence that the summit of Maunakea has always 
had within the indigenous Hawaiian community. We are most fortunate 
to have the opportunity to obtain observations from this mountain.

This work is supported by the ``Departments of Excellence 2018-2022" 
Grant awarded by the Italian Ministry of Education, University and Research 
(MIUR) (L. 232/2016). This research has made use of resources provided by 
the Compagnia di San Paolo for the grant awarded on the BLENV project 
(S1618\_L1\_MASF\_01) and by the Ministry of Education, Universities and 
Research for the grant MASF\_FFABR\_17\_01. Support for this work was provided  
by the National Aeronautics and Space Administration (NASA) grants 
GO4-15096X, AR6-17012X, and GO6-17081X. A.\ R.\ A.\ thanks the Conselho Nacional 
de Desenvolvimento Cient\'ifico e Tecnol\'ogico (CNPq) for partial support for this work.\\




\bigskip

\begin{thebibliography}{}

\bibitem[Amodeo et al.(2017)]{amodeo2017} Amodeo, S., Mei, S., Stanford, S.\ A., et al.\ 2017, ApJ, 844, 101

\bibitem[Baldi et al.(2013)]{baldi2013} Baldi, R.\ D., Capetti, A., Buttiglione, S., Chiaberge, M. \& Celotti, A. 2013, A\&A, 560, 81
	
\bibitem[Balmaverde et al.(2012)]{balmaverde2012} Balmaverde, B., Capetti, A., Grandi, P., Torresi, E., et al.\ 2012, A\&A, 545, 143
	
\bibitem[Begelman et al.(1979)]{begelman1979} Begelman, M.\ C., Rees, M.\ J. \& Blandford, R.\ D. 1979, Nature, 279, 770
								
\bibitem[Bennett et al.(1962a)]{bennett1962a} Bennett, A.\ S. 1962a, MmRAS, 68, 163

\bibitem[Bennett et al.(1962b)]{bennett1962b} Bennett, A.\ S. 1962b, MNRAS, 125, 75

\bibitem[Blanton et al.(2000)]{blanton2000} Blanton, E., Gregg, M.\ D., Helfand, D.\ J., et al.\ 2000, ApJ, 531, 118

\bibitem[Blanton et al.(2003)]{blanton2003} Blanton, E., Gregg, M.\ D., Helfand, D.\ J., et al.\ 2003, AJ, 125, 1635

\bibitem[Buttiglione et al.(2010)]{buttiglione2010} Buttiglione, S., Capetti, A., Celotti, A., et al. 2010, A\&A 509, 6
    
\bibitem[Beers et al.(1990)]{beers1990} Beers, T.C., Flynn, K. \& Gebhardt, K. 1990, AJ, 100,32

\bibitem[Campanelli et al.(2007)]{campanelli2007} Campanelli, M., Lousto, C., Zlochower, Y., Merritt, D. 2007, ApJL, 659, L5

\bibitem[Caon et al.(1993)]{caon1993} Caon, N., Capaccioli, M., \& D'Onofrio, M. 1993, MNRAS, 265, 1013


\bibitem[Cleary et al.(2007)]{cleary2007} Cleary, K., Lawrence, C.\ R., Marshall, J.\ A., et al.\ 2007, ApJ, 660, 117 

\bibitem[Coenda et al.(2005)]{coenda2005} Coenda, V., Donzelli, C.J., Muriel, H., Quintana, H., Infante, L. 2005, AJ, 129, 1237

\bibitem[Contini \& Viegas(1992)]{contini1992} Contini, M. \& Viegas, S.\ M. 1992, ApJ, 401, 481

\bibitem[de Koff et al.(1996)]{dekoff1996} de Koff, S., Baum, S.\ A., Sparks, W.\ B., et al. 1996, ApJS, 107,621

\bibitem[Donzelli et al.(2011)]{donzelli2011} Donzelli, C.J., Muriel, H., \& Madrid, J.\ P. 2011, ApJ, 195, 15

\bibitem[Edge et al.(1959)]{edge1959} Edge, D.\ O., Shakeshaft, J.\ R., McAdam, W.\ B., Baldwin, J.\ E., \& Archer, S. 1959, Mem.\ R.\ Astron.\ Soc., 68, 37

\bibitem[Evans et al.(2006)]{evans2006} Evans, D.\ A., Worrall, D.\ M., Hardcastle, M.\ J., Kraft, R.\ P., Birkinshaw, M. 2006, ApJ, 642, 96

\bibitem[Freeman et al.(1970)]{freeman1970} Freeman, K. C. 1970, ApJ, 160, 811

\bibitem[Fruscione et al.(2006)]{fruscione2006} Fruscione, A., McDowell, J.\ C., Allen, G.\ E., et al. 2006, Proc.\ SPIE, 6270, 62701V

\bibitem[Gisler \& Miley(1979)]{gisler1979} Gisler G.\ R. \& Miley, G.\ K. 1979, A\&A, 76, 109

\bibitem[Gonzalez et al.(2005)]{gonzalez2005} Gonzalez, A.\ H., Zabludoff, A.\ I., \& Zaritsky, D. 2005, ApJ,618, 195

\bibitem[Gunn et al.(1981)]{gunn1981} Gunn, J.\ E., Hoessel, J.\ G., Westphal, J.\ A., Perryman, M.\ A.\ C. \& Longair, M. S. 1981, MNRAS, 194, 111

\bibitem[Hardcastle \& Worrall(2000)]{hardcastle2000} Hardcastle, M. J. \& Worrall, D. M. 2000, MNRAS, 319, 562


\bibitem[Inskip et al.(2010)]{inskip2010}Inskip, K.\ J., Tadhunter, C.\ N., Morganti, R., et al.\  2010, MNRAS, 407, 1739

\bibitem[Jedrzejewski(1987)]{jed1987} Jedrzejewski, R. 1987, MNRAS, 226, 747

\bibitem[Komossa et al.(2008)]{komossa2008} Komossa, S., Xu, D., Zhou, H., Storchi-Bergmann, T., Binette, L. 2008, ApJ, 680, 926

\bibitem[Komossa et al.(2012)]{komossa2012} Komossa, S. 2012, Advances in Astronomy, 2012, 364973

\bibitem[Kristian et al.(1974)]{kristian1974} Kristian, J., Sandage, A. \& Katem, B. 1974, ApJ, 191, 43

\bibitem[Laing et al.(1983)]{laing1983} Laing, R.\ A., Riley, J.\ M. \& Longair, M. S. 1983, MNRAS, 204, 151 

\bibitem[Leipski et al.(2010)]{leipski2010} Leipski, C., Haas, M., Willner, S.\ P.\ et al.\ 2010, ApJ, 717, 766

\bibitem[Lynden-Bell(1969)]{lynden1969} Lynden-Bell, D. 1969, Nature, 223, 690 

\bibitem[Madrid et al.(2006)]{madrid2006} Madrid, J.\ P., Chiaberge, M., Floyd, D. et al.\ 2006, ApJS, 164, 307 

\bibitem[Madrid \& Donzelli(2013)]{madrid2013} Madrid, J.\ P. \& Donzelli, C.\ J. 2013, ApJ, 770, 158 

\bibitem[Madrid \& Donzelli(2016)]{madrid2016} Madrid, J.\ P. \& Donzelli, C.\ J. 2016, ApJ, 819, 50 

\bibitem[Martel et al.(1999)]{martel1999} Martel, A.\ R., Baum, S.\ A., Sparks, W.\ et al.\ 1999, ApJS, 122, 81

\bibitem[Marziani et al.(2016)]{marziani2016} Marziani, P., Sulentic, J.\ W., Stirpe, G.\ M., et al.\ 2016, Astrophysics and Space Science, 361, 3

\bibitem[Maselli et al.(2016)]{maselli2016} Maselli, A., Massaro, F., Cusumano, G., et al.\ 2016, MNRAS, 460, 3829

\bibitem[Massaro et al.(2009)]{massaro2009} Massaro, F., Harris, D.\ E., Chiaberge, M., et al.\ 2009, ApJ, 696, 980

\bibitem[Massaro et al.(2010)]{massaro2010} Massaro, F., Harris, D.\ E., Tremblay, G.\ R. et al.\ 2010, ApJ, 714, 589

\bibitem[Massaro et al.(2012)]{massaro2012} Massaro, F., Tremblay, G.\ R., Harris, D.\ E., et al.\ 2012, ApJS, 203, 31

\bibitem[Massaro et al.(2013)]{massaro2013} Massaro, F., Harris, D.\ E., Tremblay, G.\ R., et al.\ 2013, ApJS, 206, 7

\bibitem[Massaro et al.(2015)]{massaro2015} Massaro, F., Harris, D.\ E., Liuzzo, E., et al.\ 2015, ApJS, 220, 5

\bibitem[Massaro et al.(2018)]{massaro2018} Massaro, F., Missaglia, V., Stuardi, C., et al.\ 2018, ApJS, 234, 7

\bibitem[McCarthy et al.(1997)]{mccarthy1997} McCarthy, P., Miley, G.\ K., de Koff, S. et al. 1997, ApJS, 112, 415

\bibitem[Morganti et al.(1999)]{morganti1999} Morganti, R., Oosterloo, T., Tadhunter, C. N., et al.\ 1999, Astronomy \& Astrophysics Supplement, 140, 355

\bibitem[Morganti et al.(1993)]{morganti1993} Morganti, R., Killeen, N.\ E.\ B., Tadhunter, C. N. \ 1993, MNRAS, 263, 1023

\bibitem[Muriel et al.(2015)]{muriel2015} Muriel, H., Donzelli, C.\ J., Rovero, A.\ C.\ \& Pichel, A. 2015, A\&A, 574, A101

\bibitem[Norris et al.(2013)]{norris2013} Norris, R.\ P., Hopkins, A.\ M., Afonso, J., et al. 2013, PASA, 30, 20

\bibitem[Nurmi et al.(2013)]{nurmi2013} Nurmi, P., Hein{\"a}m{\"a}ki, P., Sepp, T., et al. 2013, MNRAS, 436 380

\bibitem[O'Dea \& Owen(1986)]{odea1986} O'Dea, C.\ P. \& Owen, F.\ N. 1986, ApJ, 301, 841

\bibitem[Osterbrock(1989)]{osterbrock1989} Osterbrock D. E. 1989 The Astrophysics of Gaseous Nebulae and Active Galactic Nuclei (Mill Valley, CA: Univ. Science Books)

\bibitem[Paterno-Mahler et al.(2017)]{paterno2017} Paterno-Mahler, R., Blanton, E.\ L., Brodwin, M., et al. 2017, ApJ, 844, 78

\bibitem[Planck Collaboration(2016)]{planck2016} Planck Collaboration XIII, 2016, A\&A, 594, 13

\bibitem[Pogge \& Owen(1993)]{pogge1993} Pogge, R. W., \& Owen, J. M. 1993, LINER; An Interactive Spectral Line Analysis Program, OSU Internal Report 93-01

\bibitem[Popovic(2012)]{popovic2012} Popovic, L.~{\v C}. 2012, New Astr. Rev. 56, 74

\bibitem[Ramos Almeida(2011)]{ramos2011} Ramos Almeida, C., Tadhunter, C.\ N., Inskip, K.\ J., et al. 2011, MNRAS, 410, 1550

\bibitem[Rovero et al.(2016)]{rovero2016} Rovero, A.C., Muriel, H., Donzelli, C.J. \& Pichel, A. 2016, A\&A, 589, 92

\bibitem[Salpeter(1964)]{salpeter1964} Salpeter, E.\ E.\ 1964, ApJ, 140, 796

\bibitem[Schlafly \& Finkbeiner(2011)]{schlafy2011} Schlafly, E., \& Finkbeiner, D. P. 2011, ApJ, 737, 103

\bibitem[Schmidt(1965)]{schmidt1965} Schmidt, M. 1965, ApJ, 141, 1

\bibitem[Seigar al.(2007)]{seigar2007} Seigar, M.\ S., Graham, A.\ W., \& Jerjen, H. 2007, MNRAS, 378, 1575

\bibitem[S\'ersic(1968)]{sersic1968} S\'ersic, J.\ L. 1968, Atlas de Galaxias Australes,
C\'ordoba, Argentina: Observatorio Astron\'omico, 1968

\bibitem[Sijbring \& Bruyn(1998)]{sijbring1998} Sijbring, D. \& de Bruyn, A.\ G. 1998, A\&A, 331, 901


\bibitem[Smith \& Spinrad(1988)]{smith1980} Smith, H.\ E.\ \& Spinrad, H. 1980, PASP, 92, 553

\bibitem[Spinrad \& Djorgovski(1984)]{spinrad1984} Spinrad, H. \&  Djorgovski, S. 1984, ApJ Letters, 285, L49

\bibitem[Spinrad et al.(1985)]{spinrad1985} Spinrad, H., Marr, J., \& Aguilar, L., Djorgovski, S. 1985, PASP, 97, 932

\bibitem[Stuardi et al.(2018)]{stuardi2018} Stuardi, C., Missaglia, V., Massaro, F., Ricci, F., Liuzzo, E., 2018, ApJS, 235, 32

\bibitem[Tadhunter et al.(1993)]{tadhunter1993} Tadhunter, C.\ N., Morganti, R., di Serego-Alighieri, S., et al.\ 1993, MNRAS, 263, 999

\bibitem[Tremblay et al.(2009)]{tremblay2009} Tremblay, G.\ T., Chiaberge, M., Sparks, W. B. et al. 2009, ApJS, 183, 278

\bibitem[Westhues et al.(2016)]{westhues2016} Westhues, C., Haas, M., Barthel, P.\ et al.\ 2016, AJ, 151, 120

\bibitem[Wilkes et al.(2013)]{wilkes2013} Wilkes, B. J., Kuraszkiewicz, J., Haas, M., et al.\ 2013, ApJ, 773, 15

\bibitem[Wing \& Blanton(2011)]{wing2011} Wing, J.\ D. \& Blanton, E. L. 2011, AJ, 141, 88

\bibitem[Wright(2006)]{wright2006} Wright, E. L. 2006, PASP, 118, 1711 

\bibitem[Wyndham et al.(1966)]{wyndham1966} Wyndham, J.\ D. 1966, ApJ, 144, 459 

\bibitem[Zakamska et al.(2016)]{zakamska2016} Zakamska, N. L., Hamann, F., P\^aris, I., et al. 2016, MNRAS, 459, 3144

\end{thebibliography}
\end{document}